\documentclass[aps,pre,a4paper,onecolumn,notitlepage,nofootinbib]{revtex4-1}
\usepackage[shortlabels]{enumitem}
\usepackage{graphicx}
\usepackage{subfig}

\usepackage{amsmath}
\usepackage{hyperref}
\usepackage{url}
\usepackage{amssymb}
\usepackage{amsfonts}
\usepackage{indentfirst}
\usepackage[toc,title]{appendix}
\usepackage{color}

\usepackage{natbib}

\newcommand{\<}{\left<}
\renewcommand{\>}{\right>}
\newcommand{\idm}{\mathbf{1}}

\newcommand{\cH}{{\cal H}}

\newcommand{\ket}[1]{\left| #1 \>}
\newcommand{\bra}[1]{\< #1 \right|}
\newcommand{\braket}[2]{\< #1 | #2 \>}

\newcommand{\be}{\begin{eqnarray}}
\newcommand{\ee}{\end{eqnarray}}

\newcommand{\MeijerG}[8][\bigg]{G^{{ #2 },{ #3 }}_{{ #4 },{ #5 }} #1( \begin{matrix} #6 \\ #7 \end{matrix}\, #1\vert\, #8 #1)}

\begin{document}

\title{Narain transform for spectral deformations of random matrix models } 
\author{M. A. Nowak\thanks{maciej.a.nowak@uj.edu.pl} }
\email{maciej.a.nowak@uj.edu.pl}
\author{W.  Tarnowski\thanks{wojciech.tarnowski@doctoral.uj.edu.pl}}
\email{wojciech.tarnowski@doctoral.uj.edu.pl}
\affiliation{Marian  Smoluchowski Institute of Physics and
Mark Kac Complex Systems Research Center, Jagiellonian University, S. \L{}ojasiewicza 11, PL 30-348 Krak\'ow, Poland.}

\date{September 13, 2019}

\begin{abstract}
We start from applying   the general idea of  spectral projection (suggested by Olshanski and Borodin and advocated by Tao)  to the complex Wishart model.  Combining the ideas of spectral projection with the insights from quantum mechanics we derive in an effortless way all  spectral properties of the complex Wishart model:
first,   the Marcenko-Pastur distribution interpreted as a Bohr-Sommerfeld quantization condition for the hydrogen atom; second,  hard (Bessel), soft (Airy)  and bulk (sine) microscopic kernels from properly rescaled  radial Schr\"odinger equation for the hydrogen atom.  Then, generalizing the ideas based on Schr\"odinger equation to  the case when Hamiltonian is non-Hermitian,  we  propose an analogous construction  for spectral projections of universal kernels built from bi-orthogonal ensembles. In particular, we  demonstrate that the Narain transform is a natural extension of the Hankel transform for the products of Wishart matrices, yielding an explicit form of the universal kernel at the hard edge.  We also show, how the change of  variables of the {\it rescaled} kernel allows to make the link to  universal kernel of  Muttalib-Borodin ensemble. The proposed construction  offers a simple alternative  to standard methods of derivation of  microscopic kernels,  based e.g. on 
Plancherel Rotach limit of orthogonal polynomials of  asymptotics in the Riemann-Hilbert  problem. Finally, we speculate, that a suitable  extension of the Bochner theorem for Sturm-Liouville operators may provide an additional insight into the classification of
microscopic universality classes in random matrix theory.

\end{abstract}
\maketitle

\section{Introduction}

Determinantal point processes~\cite{Determinantal} appear in several  areas of mathematics, physics and applied sciences,  ranging from random matrix theory to combinatorics  and theory of  representations. The unique feature of such processes relies on the fact, that the $N$-point joint probability distribution function is expressed as a determinant built from a single, two-point correlation function known as a kernel. 
Celebrated examples  of such kernels in high energy physics include ``hard-edge" Bessel kernels~\cite{Shuryak1993,Verbaarschot1994a,Verbaarschot1994b}, observed in numerous lattice calculations~\cite{HELLER},  
or Pearcey kernels appearing at strong-weak coupling phase transition in  Yang-Mills theories in the limit of large number of colors~\cite{Narayanan2007,Lohmayer2009}.
The calculation of kernels  and their asymptotic limits became therefore an area of vigorous studies using advanced mathematical tools, like  supersymmetry~\cite{Efetov1983,Verbaarschot1985a,Verbaarschot1985,Guhr1991}, orthogonal~\cite{MEHTA} and bi-orthogonal polynomials~\cite{BorodinBiOrtho}, Riemann-Hilbert problem~\cite{Bleher1999,Deift1998,Deift1999} and Plancherel-Rotach~\cite{Plancherel1929} limiting procedures for integral representations, to mention most popular. 

 Borodin and Olshanski~\cite{Borodin2007} offered a different point of view at kernels in random matrix theory built from orthogonal polynomials.  When treated as an integral operator, the kernel is a projection -- a consequence of a finite number of eigenvalues and orthogonality of polynomials.  This idea was later advocated by Tao~\cite{TAO}, who also gave physical intuition using the mapping between Gaussian Unitary Ensemble and the quantum harmonic oscillator. In this quantum mechanical picture the projection stems from the fact that the first $N$ energy levels are occupied. Using these techniques, Bornemann elaborated the Sturm-Liouville problem and showed that all three classical limiting kernels can be obtained in this way~\cite{BORNEMANN}.

The aim of this work is to further elaborate the spectral projection  method, with the use of insights from elementary quantum mechanics. In section 2, we pedagogically introduce the spectral projection method and demonstrate its easiness in taking the microscopic limits by  recalculating all limiting kernels in the complex Wishart ensemble.
% {\color{red} In particular, we rederive the microscopic hard edge kernel which gives the microscopic spectral density of the lattice Dirac operator~\cite{Verbaarschot1994,Jackson1996,Berbenni1998,Wettig1997} }
We link the Marchenko-Pastur distribution to the Bohr-Sommerfeld quantization condition. 
 We also notice that the threeness  of  the classical universal kernels  can be linked to the strictures originating from the Bochner theorem  for Sturm-Liouville problem~\cite{Bochner}.
 
Recent developments on the integrable structure of products of random matrices and the multitude of new microscopic kernels in biorthogonal ensembles naturally pose a question whether the spectral projection method can be extended to incorporate these universality classes. In section~3 we discuss the possibilities to circumvent the constrains of Bochner's theorem  and consider an analog of a quantum-mechanical Hamiltonian, but  with higher number of derivatives. Although such an operator may not be self-adjoint, still, due the fact that its left and right eigenvectors form a bi-orthogonal basis, it is possible to infer the microscopic limit of the kernels using the spectral projection method. We demonstrate this on two examples -- singular values of products of Gaussian matrices~\cite{AIKWishart} and the Muttalib-Borodin ensemble~\cite{Muttalib,Borodin}. In both cases the Narain transform~\cite{Narain1,Narain2,Narain3} allows one to recover the Meijer-G hard edge universality, generalizing Bessel kernel. Again, the spectral projection translates to the truncation of the phase space of the associated transform.

Section~4  concludes the paper.  In appendix~\ref{sec:Hydro2D}, we show an alternative mapping of the Wishart ensemble to the 2-dimensional hydrogen atom problem~\cite{Hydrogen2D}.  In appendix~\ref{sec:AppWKB} we recover the Marchenko-Pastur distribution from the WKB approximation.  In appendix~\ref{sec:MeijerG} we recall some properties of the Meijer-G functions.

\section{Spectral projections from hydrogen atom problem}
\subsection{Complex Wishart ensemble} 
Let us consider Hermitian matrix  $M=XX^{\dagger}$, where $X$ is the complex $N \times T$ matrix with entries given by the probability density function $P(X)dX=Z_{NT}^{-1} e^{-\frac{1}{\sigma^2} \sum_{\alpha,j}^{N,T}|X_{\alpha, j}|^2} \prod_{\alpha j}^{N,T} d\Re X_{\alpha j} d\Im X_{\alpha j}$. Here $Z_{NT}^{-1}$ provides the normalization and $\sigma^2$ is the variance of the complex Gaussian distribution, which we set to 1, to simplify the expressions.  This defines complex Wishart matrix~\cite{WISHART}. Switching to eigenvalues, we arrive, using standard methods~\cite{PASTUR}, at their joint probability density
\be 
P(\lambda_1, ...,\lambda_N)=Q_N^{-1}\prod_{j=1}^N \lambda_j^{\alpha} e^{-\lambda_j} \prod_{1 \le i <j \le N} |\lambda_i-\lambda_j|^2,
\ee 
with $\alpha=T-N$, and  the Vandermonde  determinant (last  term) is the price for switching from elements of $X$ to eigenvalues $\lambda_i$ of matrix $M$. Standard orthogonal polynomials trick~\cite{MEHTA} allows one to rewrite the probability distribution as 
\be
P_N(\lambda_1,...,\lambda_N)=\frac{1}{N!} \left( \det \left[\psi_{j-1}(\lambda_k)\right]_{j,k=1}^N \right)^2= \frac{1}{N!} \left[\det K_N(\lambda_i, \lambda_j)\right],
\label{Slater}
\ee
with the correlation {\em kernel} 
\be
K_N(\lambda,\mu)=\sum_{l=0}^{N-1}\psi_l(\lambda)\psi_l(\mu),
\ee
where $\psi_l(\lambda)=  e^{- \lambda/2} \lambda^{\alpha/2} P_l(\lambda)$ and $P_l$ are monic polynomials. This form already suggests  links to quantum mechanics. The first equality in (\ref{Slater}) represents the joint probability of eigenvalues as the square of the Slater determinant, therefore can be interpreted as the quantum probability density of non-interacting spinless fermions (see~\cite{Fermions} for a review). This also explains why the eigenvalue density is expressed solely in terms of a two-point function (second expression on the r.h.s. of (\ref{Slater})). Next, we see that the most natural choice of polynomials  is dictated by the weight $w_{\alpha}(\lambda)=\lambda^{\alpha}e^{-\lambda} $. 
Such polynomials,  orthonormal on the positive part of the real axis,  are the associated Laguerre polynomials and appear in the radial part of the Schr\"odinger equation. 
Indeed, upon standard separation of variables in the wavefunction, $\varphi(\vec{r})=R(r)Y^{m}_l(\theta,\psi)$, it reads 
\be
\frac{d^2 y(r)}{dr^2} +\left[\frac{2\mu e^2}{r \hbar^2}  - \frac{l(l+1)}{r^2}\right]y(r)=-\frac{2\mu E}{\hbar^2}y(r),
\label{Sch}
\ee
where $y(r)=rR(r)$.  Switching to dimensionless variable $x=r \epsilon$, where
$(\epsilon/2)^2=-2\mu E/\hbar^2$, putting $2\mu=1$ and all other physical constants to 1, we recover~\cite{WEINBERG}
\be
\frac{d^2 y(x)}{dx^2} +\left[-\frac{1}{4} +\frac{1}{\epsilon x} - \frac{l(l+1)}{x^2}\right]y(x)=0,
\label{Schreps}
\ee
where $y=y^l_n= e^{-x/2}x^{(k+1)/2}L^k_j(x)$. Here $k=2l+1$ and the principal quantum number is related to the order of Laguerre polynomial as $n=j+l+1$.  Note,  that $\epsilon=1/n$, or, equivalently, $E_n=-1/4n^2$, since in our units Bohr's radius equals to 2.   To map this random matrix problem to the hydrogen atom we associate $\psi_l(\lambda)=\sqrt{x}y(x) $. 
This completes the dictionary between  hydrogen atom problem and the Wishart kernel. In Appendix \ref{sec:Hydro2D} we also present a mapping into 2D hydrogen atom with $1/r$ potential~\cite{Hydrogen2D}, in which the relation between eigenfunctions of the radial part of the Schr\"odinger equation and $\psi$ is even more explicit.

The Schr\"odinger equation for $\psi$ expressed in terms of the parameters of the Wishart ensemble reads
\be
\frac{d^2 \psi_k}{dx^2} + \frac{1}{x} \frac{d\psi_k}{dx} +\frac{1+2k+\alpha}{2x} \psi_k -\frac{\alpha^2}{4x^2}\psi_k =\frac{1}{4} \psi_k. \label{mainequation_no1}
\label{MAIN}
\ee
 
 Finally, let us note that   in the bra-ket notations the kernel reads 
  $\hat{K}_N=\sum_{k=0}^{N-1}\ket{\psi_k}\bra{\psi_k}$ thus it is the operator projecting onto the set of $N$ lowest eigenstates. Indeed, due to the orthonormality  of eigenfunctions $\hat{K}_N^2=\hat{K}_N$. Last equation, when calculated in coordinate  representation,  yields well-known reproducing property $\bra{x}\hat{K}_N\ket{y} \equiv K_N(x,y)= \int K_N(x,z)K_N(z,y)dz$.

\subsection{Macroscopic  density from the semiclassical approximation}
To have the finite support of the spectral density in the large $N$ limit, we rescale  $x\to Tx$. Upon this scaling and identifying momentum\footnote{For this analogy it is even better to take the 2D radial momentum $p_r=\frac{i}{T}\left(\frac{d}{dx}+\frac{1}{x}\right)$, but this eventually leads to the same result in large $T$ limit.} as $p=-\frac{i}{T}\frac{d}{dx}$ (in analogy to $\hbar \leftrightarrow 1/T$) in the limit $N,T\to\infty$ with $c=N/T$ fixed we obtain the Schr\"odeinger equation $(p^2+V_{eff})\psi=-\frac{1}{4}\psi$ with the effective potential
\begin{equation}
V_{eff}=\frac{(1-c)^2}{4x^2}-\frac{1+c}{2x}.
\end{equation}

Thanks to the fermionic analogy, the mean spectral density is the same as the probability density of non-interacting fermions. The latter is obtained by integrating the Wigner function over the set of momenta. In our case the Wigner function is constant on the region of the phase space $p^2+V(x)\leq \frac{1}{4}$ and zero outside~\cite{WignerFunction} (see Fig.~\ref{Fig:MP}), therefore the density of eigenvalues is proportional to the momentum and the Bohr-Sommerfeld quantization condition 
\be
T\oint p(xT)  dx= \left(N+\frac{1}{2}\right) 2\pi 
\ee
on the RMT side corresponds to the normalization of density
\be
\int_{r_-}^{r_+} \rho(x) dx =1.
\ee
This allows us to obtain the density of eigenvalues 
\be
\rho(x)= \frac{1}{2\pi c x} \sqrt{(r_+ - x)(x-r_-)}.
\label{MP}
\ee
Here $r_{\pm}=(1 \pm \sqrt{c})^2$ are classical turning points in WKB approximation. In Appendix~\ref{sec:AppWKB} we provide another derivation of this result based on the explicit WKB analysis of \eqref{mainequation_no1}.

We have therefore obtained Marchenko-Pastur distribution as an {\it exact, semiclassical limit  of the quantum mechanical hydrogen  atom problem}. It is intriguing to speculate  why such link has not  been exploited (to the authors knowledge) in the literature. 
Perhaps the reason  is that Bohr-Sommerfeld quantization condition does not reproduce correctly the ground state of the hydrogen atom, and not even the Bohr quantization condition~\cite{NOTE}.
 It is amusing to notice, that if one replaced
$l(l+1)$ by $(l+1/2)^2$ in the numerator of the centrifugal potential, this would be the case and B-S approximation would lead to the exact result for the hydrogen spectra~\cite{LANGER}. Of course, in the large $l$ limit it does not matter which of the equations  (\ref{Schreps}) or ({\ref{mainequation_no1}) we use, however, at the microscopic level, additional square root   in Laguerre function for the Wishart will play the crucial role in getting the proper scaling of the hard edge. 
 
We complete this part with the observation, that in the case of harmonic oscillator, similar construction is ambiguities free, since Bohr-Sommerfeld quantization condition yields exact spectrum. The Wigner semicircle, or rather semi-ellipse, is just the similar projection of the ellipse $p^2+x^2/4=1$ onto the $x$ axis in the phase space. The Bohr-Sommerfeld quantization condition just reads  $\int \rho(x)dx =1$, where $\rho=\frac{1}{2\pi}\sqrt{4-x^2}$ (in units where $2\mu=1$)~\cite{TAO}. 
Again, the rigid argument comes from the fact, that the Wigner function for harmonic oscillator is  explicitly known~\cite{WignerFunction}, and  yields a direct relation between the momenta and positions at the semi-classical level.

\label{sec:stability}

\subsection{Microscopic scaling as a spectral deformation}

\begin{figure}
\includegraphics[width=0.49\textwidth]{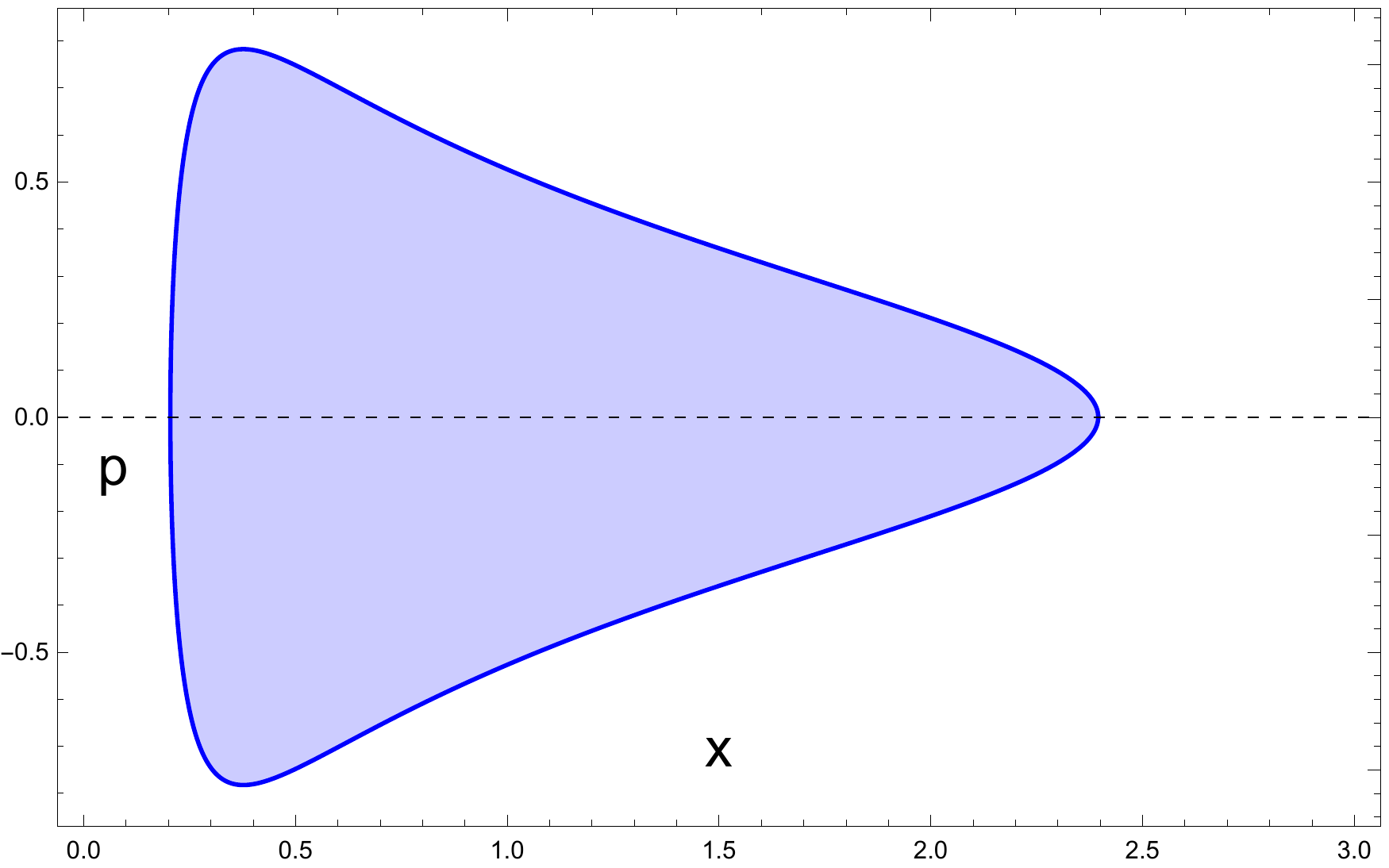}
\includegraphics[width=0.49\textwidth]{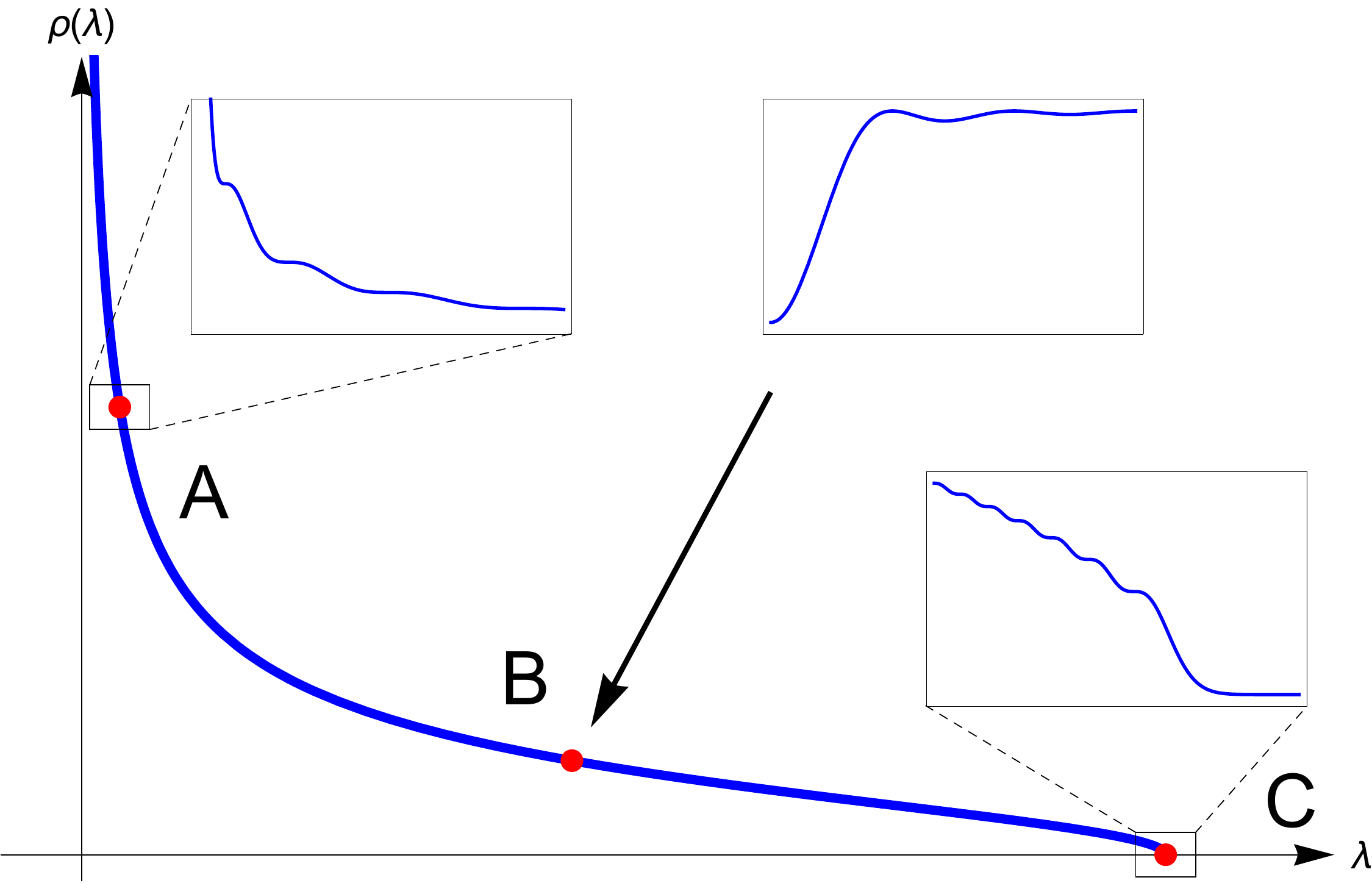}
\caption{(left) A region in the phase space where the Wigner function is nonzero. We took c=0.3. (right) Identification of three regions of the Marchenko-Pastur distribution that give rise to different microscopic scalings.
\label{Fig:MP}
}
\end{figure}

Correlations of eigenvalues probed on the scale of the typical separation between them are independent on the probability density function of matrix elements. They fall into several classes, depending on the point $x_0$ of the spectrum at which their behavior is probed. The shape of the spectral density, in turn, determines the microscopic scale $s$ by demanding that in the interval $[x_0,x_0+s]$ one expects one eigenvalue to occur.
 Looking at the form of the Marchenko-Pastur distribution (see Fig.~\ref{Fig:MP}), we immediately identify three distinct regions corresponding to microscopic scalings.

\begin{enumerate}[A]
\item{\bf{Hard edge.}}
In the limit when $N,T\to\infty$ but $\alpha=T-N$ remains fixed ($c\to 1$), the turning point $r_{-}$ approaches zero, and the eigenvalue density  near this point behaves like $1/\sqrt{x}$. Asking how many out of original $N$ eigenvalues will appear in a narrow bin of size $s$ around zero, we get
\be
n_{hard} \sim N \int_0^s \frac{dx}{\sqrt{x}} \sim N\sqrt{s}.
\ee
Demanding that $n_{hard}\sim 1$, we set the proper microscopic scale to $s \sim N^{-2}$.

\item{\bf{Bulk.}}
Between the endpoints, at some $x_0$, when counting the number of eigenvalues in a narrow interval of length $s$, one can approximate the density as locally constant $\rho(x_0)$. This leads to 
\be 
n_{bulk} \sim    N \int_{x_0-s/2}^{x_0+ s/2} \rho(x_0) dx  \sim Ns\rho(x_0),
\ee
which implies that the bulk microscopic scale is $s \sim \frac{1}{N \rho(x_0)}$.

\item{\bf{Soft edge.}}
When $c \neq1$, the macroscopic spectral density around both turning points vanishes like $\sqrt{|r_{\pm}-x|}$.
Counting the eigenvalues close to the edge, leads to
\be
n_{soft} \sim N \int_{0}^{s} \sqrt{x} dx  \sim Ns^{3/2},
\ee
thus the edge microscopic scale is set to $s \sim N^{-2/3}$.

\end{enumerate}

Following the generic arguments by Borodin and Olshanski~\cite{Borodin}  and  inspired by Tao~\cite{TAO} presentation for  the Gaussian Unitary Ensemble, we will now obtain the microscopic, universal kernels for the complex Wishart ensemble.  We remark that this case belongs to the generic class of Sturm-Liouville operators, considered recently by  Bornemann~\cite{BORNEMANN}.   However, in this note, we attempt to use the insights from quantum mechanics rather than abstract mathematics.

 The complete set of eigenfunctions provides a resolution of identity $\idm =\sum_{k=0}^{\infty}\ket{\psi_k}\bra{\psi_k}$. The random matrix kernel is obtained by truncating this sum to first $N$ eigenstates and is therefore a projection. The range of this projection can be formally written as $\hat{H}\leq E_{N-1}$
or,  using the explicit form of (\ref{MAIN}), as 
\be
\frac{d^2}{dx^2} + \frac{1}{x} \frac{d}{dx} +\frac{1+2k+\alpha}{2x}  -\frac{\alpha^2}{4x^2}  \geq \frac{1}{4}.
\label{generic}
\ee
The microscopic scalings provide further deformations of the projection range, which in the large $N,T$ limit gives rise to the universal microscopic kernels, which we work out in details beneath.

\begin{enumerate}[A]

\item{\bf{Bessel kernel.}} Using the hard edge scaling $x/T \rightarrow s N^{-2}$, and performing the large $N$ limit (note that $k \sim N$), we obtain the equation 
\be
\frac{d^2}{ds^2} +\frac{1}{s}\frac{d}{ds} +\frac{1}{s}-\frac{\alpha^2}{4s^2} \geq 0.
\ee
Changing variables $z=2\sqrt{s}$ converts the  above bound onto more familiar form 
\be
\Delta_{\alpha} \equiv -\frac{d^2}{dz^2}-\frac{1}{z}\frac{d}{dz} +\frac{\alpha^2}{z^2} \leq 1,
\label{maintrans1}
\ee
where on the l.h.s. we recognize Bessel operator, appearing in quantum mechanical problems with polar angle symmetry. To see the deformation caused by microscopic scaling at the hard edge,  we invoke the {\it  Hankel transform}
\be
F(t)=H_{\alpha}[f(z)]=\int_0^{\infty} J_{\alpha}(tz)  f(z)z dz 
\ee
 and its inverse 
 \be
 f(z)=\int_0^{\infty}J_{\alpha}(tz) F(t) t dt.
 \ee
 Since the Hankel transform of the Bessel operator reads $H_{\alpha}[\Delta_{\alpha} f(z)]=t^2 F(t)$~\cite{HANKEL}, the spectral deformation in dual variable $t$ (note that $t$ cannot be negative) reads simply 
 \be
 t\leq 1.
 \label{defHankel}
 \ee
 Hankel transform and its inverse give a representation of the identity operator 
 \be
 f(t{'})=\int_0^{\infty}\int_0^{\infty} zt J_{\alpha}(t{'}z)J_{\alpha}(tz)F_{\alpha}(t)   dt dz.
 \ee
 The deformation condition (\ref{defHankel}) restricts the range of the parameter $t$ and therefore turns  the above identity operator into the projection
 \be
 {\bf P}[f(t{'})]=\int_0^{\infty} \left[ \int_0^1 zt J_{\alpha}(t{'}z)J_{\alpha}(tz)F_{\alpha}(t)  dz \right] dz.
 \ee
 Changing variables once more as $z=\sqrt{s}$ and introducing $t=\sqrt{y}$ and $t{'}=\sqrt{x}$, we rewrite the above as 
 \begin{equation}
  {\bf P}[f(x)]=\int_0^{\infty} \left[\frac{1}{4}\int_0^{1} J_{\alpha}(\sqrt{xs})J_{\alpha}(\sqrt{sy}) ds \right]f(y)dy \equiv \int_0^{\infty} K(x,y)f(y)dy,
 \end{equation}
so the kernel, understood as a projection, reads 
\be
K_{Bessel}(x,y)=\frac{1}{4}\int_0^1 J_{\alpha}(\sqrt{xs})J_{\alpha}(\sqrt{ys})ds= 
\frac{ J_{\alpha}(\sqrt{x})  J^{\prime}_{\alpha}(\sqrt{y}) \sqrt{y} - \sqrt{x} J^{\prime}_{\alpha}(\sqrt{x})J_{\alpha}(\sqrt{y})}{2(x-y)},
\label{eq:BesselKernel}
\ee
where  on the r.h.s. we presented the more familiar form of the kernel based on the Lommel integral and primes denote differentiation with respect to the argument. Hard edge scaling deforms the upper half plane in $s$ variable onto the strip between the parallel lines $s=0$ and $s=1$.

\item {\bf{Sine kernel.}} Combining the rescaling needed for the finite support and the microsopic scaling we define the new variable $s$ as  $x/T =x_0+ \frac{s}{N\rho(x_0)}$. Upon taking the large $N,T$ limit, the bound \eqref{generic} in this new variable reads
\be
\frac{d^2}{ds^2}  \geq  \frac{(x_0-r_+)(x_0-r_-)}{4c^2x_0^2\rho^2(x_0)}.
\ee
Using  the explicit form of the Marcenko-Pastur density \eqref{MP},  the  above bound  is simplified to
\be
-\frac{d^2}{ds^2} \leq \pi^2.
\label{maintrans1}
\ee
On the l.h.s. we recognize the Schr\"odinger operator for a free particle, therefore the natural procedure for resolving this bound is to use plane waves, i.e.  move to the momentum space via the Fourier transformation:
\be
F(q) &=&\int_{-\infty}^{\infty}e^{2\pi itq}f(t) dt, \nonumber \\
f(t) &=& \int_{-\infty}^{\infty} e^{-2\pi itq}F(q)dq. 
\ee
The spectral deformation in the momentum space reads therefore 
\be q^2\leq  \frac{\pi^2}{(2\pi)^2} = \frac{1}{4}.
\label{defFour}
\ee
Combination of Fourier transforms provides a representation of an identity operator
\be
f(t^{'})=\int_{-\infty}^{\infty}\int_{-\infty}^{\infty}e^{-2\pi it'q}e^{2\pi itq}f(t) dt dq. 
\ee
The deformation (\ref{defFour}) projects the above identity operator onto 
\be
{\bf P}[f(t')] =\int_{-\infty}^{\infty} \left[  \int_{-\frac{1}{2}}^{\frac{1}{2}}e^{-2\pi it'q}e^{2\pi itq}dq \right] f(t) dt, 
\ee
Microscopic scaling in the bulk restricts the range of momenta to $-\frac{1}{2}\leq q\leq \frac{1}{2}$. Calculation of the integral in square brackets yields the projection in the position basis, which is the sine kernel
\be
K_{Sine}(t,t')=\frac{\sin(\pi(t'-t))}{\pi(t'-t)}.
\ee

 \item{\bf{Airy  kernel.}} At the soft edge we introduce the scaling variable $s$ as $x /T = r_{\pm}\pm \frac{s}{\sqrt{c} (r_{\pm}N)^{2/3}}$.  In the large $N$ and $T$ limit   generic 
 bound (\ref{generic}) is transformed into
\be
-\frac{d^2}{ds^2} +s  \leq 0.\label{eq:DiffAiry}
\label{maintrans1}
\ee
On the l.h.s. we recognize  the Schr\"odinger operator with the linear potential. This condition in the position-momentum space $(s,q)$ restricts the range of integration to the parabola $4\pi^2 q^2+s\leq 0$, which is not well suited for reading out the limiting kernel. To circumvent this problem, Tao introduced a similarity transformation in the momentum space~\cite{TAO}.   Alternatively, since we identify the differential Airy operator in \eqref{eq:DiffAiry}, we can directly  resort to \textit{the Airy transform}~\cite{VALLET}
\be
F(z)=A[f(t)]=\int_{-\infty}^{\infty} Ai(z-t)f(t)dt
\ee
 and its inverse
 \be
 f(t)=\int_{\infty}^{\infty}F(z)Ai(z-t)dz.
 \ee
 Using the Airy transform  for the operator bound  \eqref{eq:DiffAiry}, and the fact that Airy function fulfills $Ai{''}(x)=xAi(x)$ we express the spectral deformation in dual variable $t$ simply as 
 \be z\leq 0.
 \label{defAiry}
 \ee
 Combining  both Airy transforms we obtain  the identity operator
 \be
 f(t{'})=\int_{-\infty}^{\infty}\int_{-\infty}^{\infty}Ai(t{'}-z)Ai(t-z)f(t)dtdz
 \ee
 The deformation condition (\ref{defAiry}) turns the above identity operator into a projection 
\be
{\bf P}[f(t{'})]=\int_{-\infty}^{\infty} \left[ \int_{-\infty}^0 Ai(t{'}-z)Ai(t-z) dz \right] f(t) dt
\ee
so the kernel reads 
\be
K_{Airy}(t,t{'})= \int_{-\infty}^0  Ai(t{'}-z)Ai(t-z) dz=\frac{Ai(t{'})Ai{'}(t)-Ai{'}(t{'})Ai(t)   }{t{'}-t},
\ee 
where on the r.h.s. we presented the more familiar form of the Airy kernel  based on relation
\be
\frac{d}{dz}\left[   \frac{   Ai(t{'}-z)Ai{'}(t-z)-Ai{'}(t{'}-z)Ai(t-z)       }{t{'}-t}        \right]=
Ai(t{'}-z)Ai(t-z).
\ee

\end{enumerate}

We summarize  this section in Fig.~\ref{Fig:deform}, by plotting the domain of the projection operator before and after the pertinent microscopic scalings.  

\begin{figure}[ht!]
      \includegraphics[ width=0.3\textwidth]{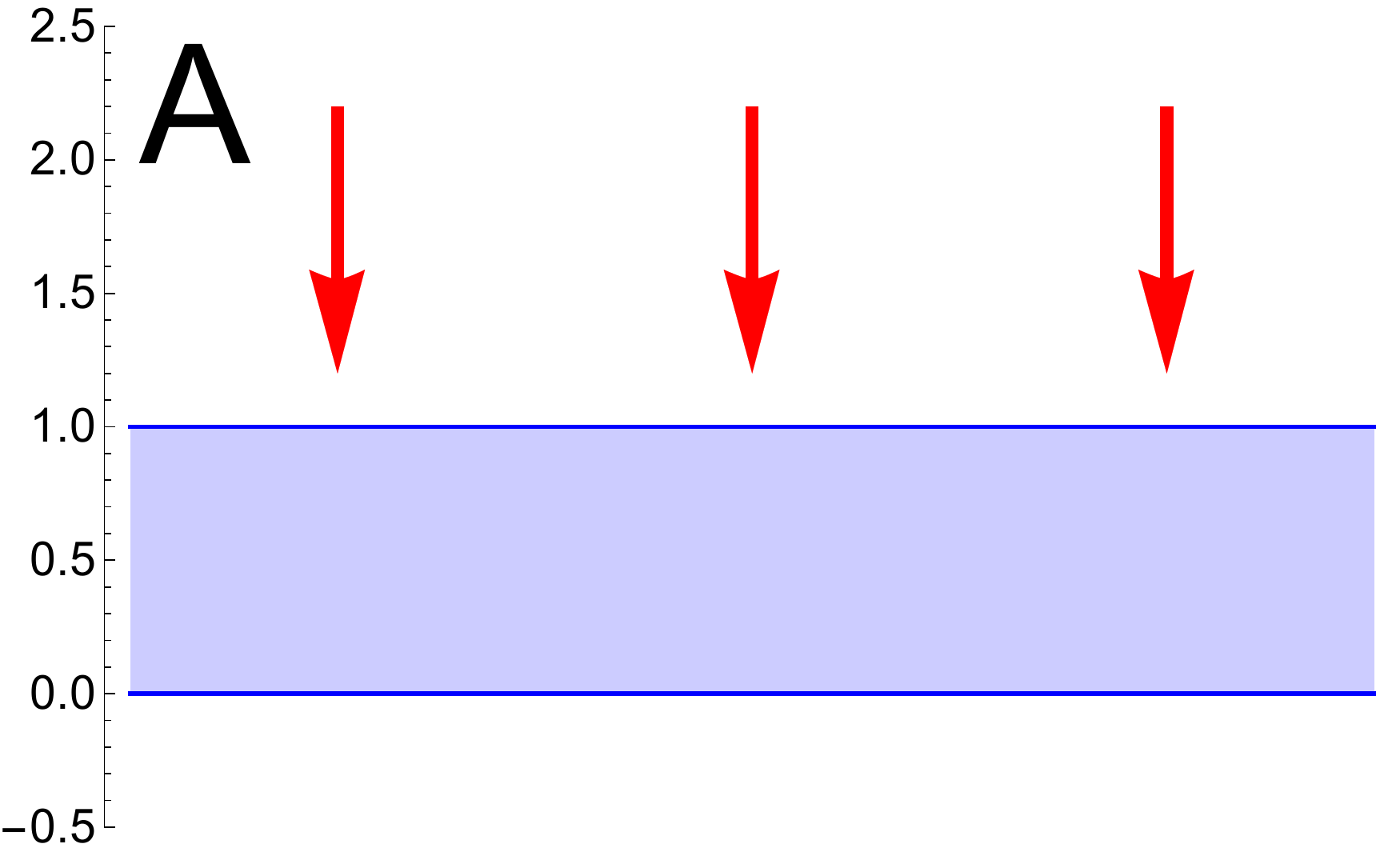}
      \includegraphics[ width=0.3\textwidth]{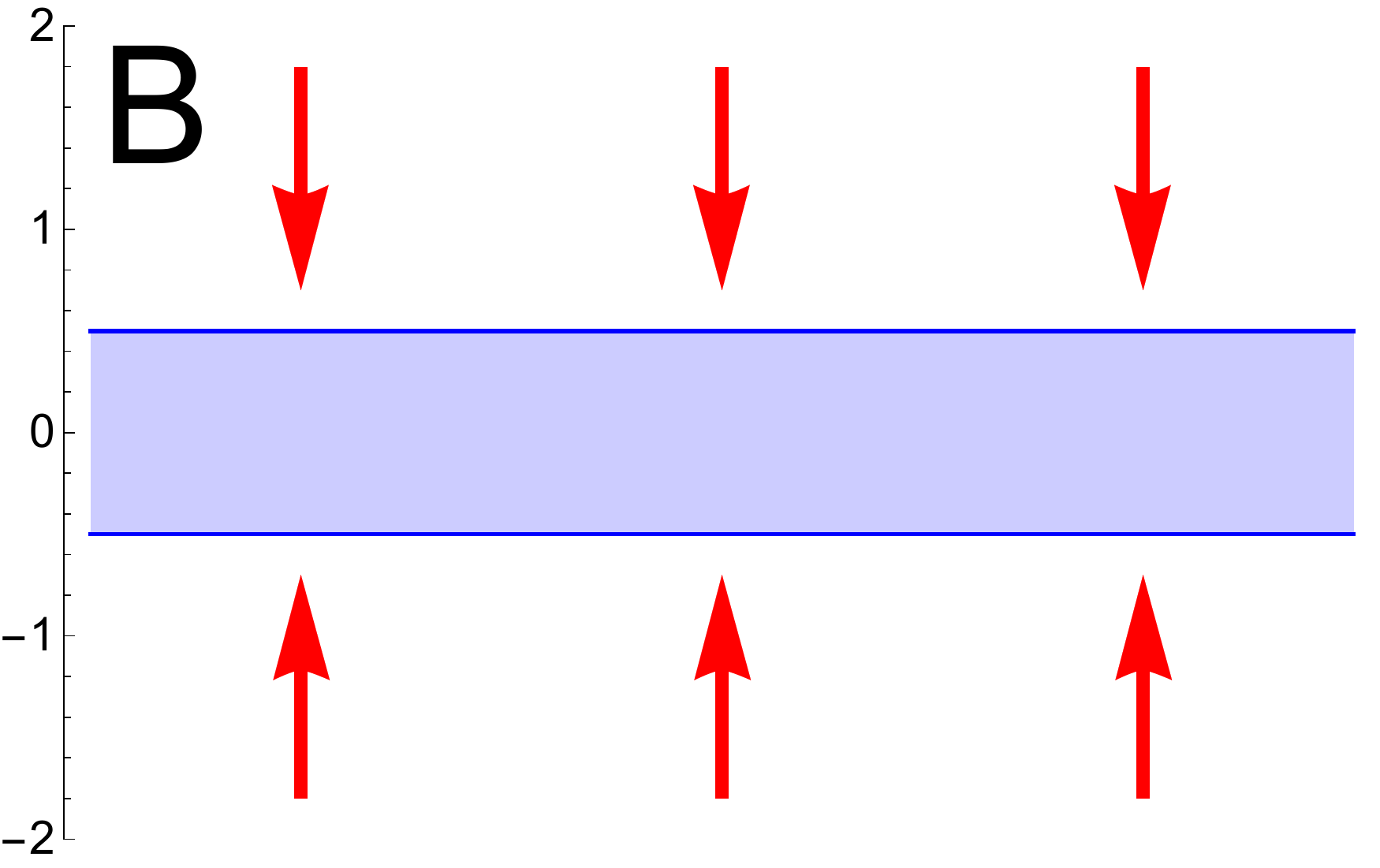}
      \includegraphics[ width=0.3\textwidth]{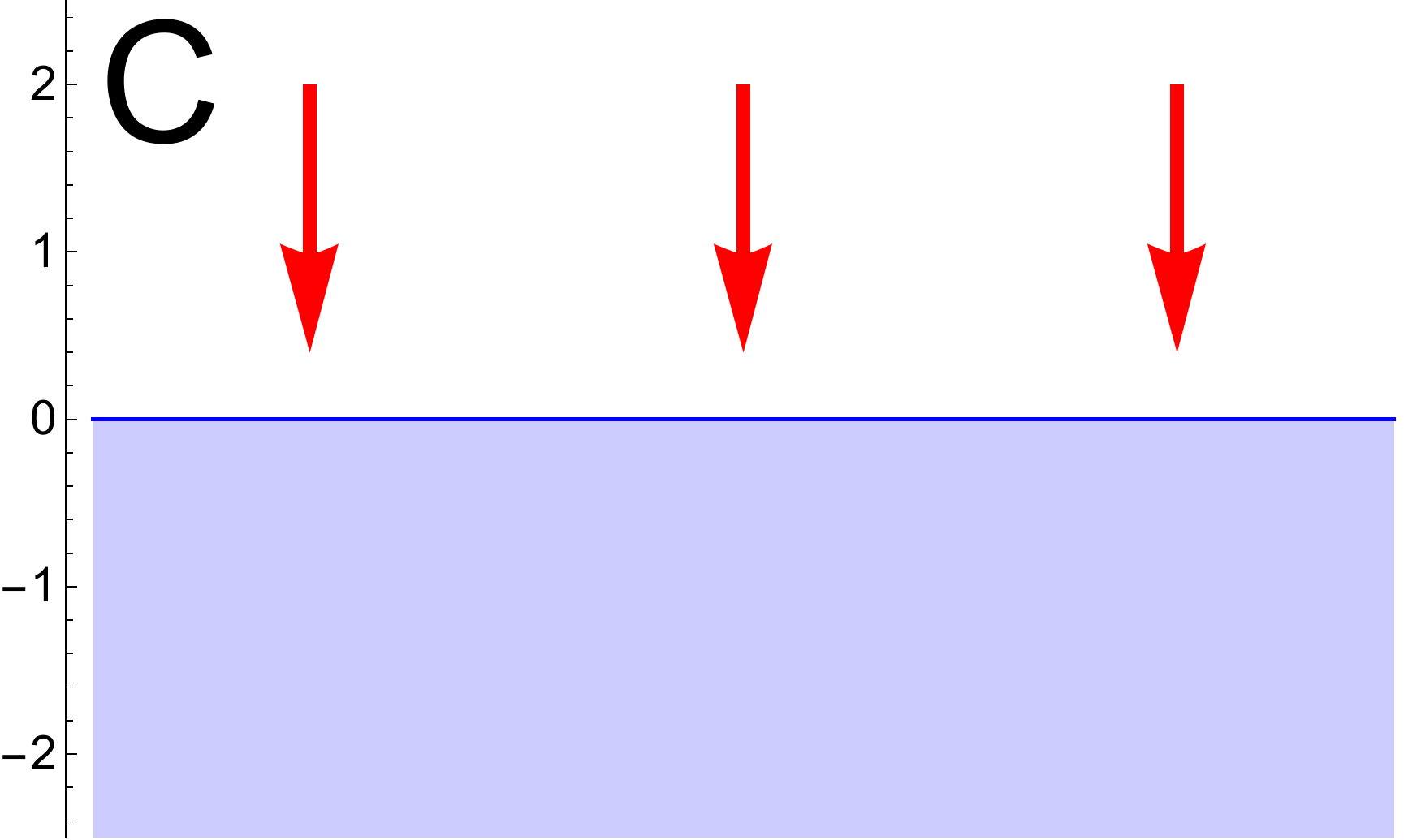}
\caption{ Regions in the phase space after microscopic scaling  at the hard edge (A), 
in bulk (B) and at the soft edge (C).  Red arrows point at the direction of deformation. 
\label{Fig:deform}}
\end{figure}

\section{Bochner theorem and beyond -  non-Hermitian Hamiltonians.}

\subsection{Bochner theorem} 
In a short paper written in 1929~\cite{Bochner}, Salomon Bochner has noticed,  that if an infinite  sequence of polynomials $P_n(x) $ satisfies an eigenequation to the second order self-adjoint differential operator
\be
p(x)P^{\prime\prime}_n(x) +q(x) P_n^{\prime}(x) +r(x)P_n(x)=\lambda_n P_n(x),
\ee
then $p(x), q(x), r(x)$ must be polynomials of  degree 2, 1, and 0, respectively.  If additionally  polynomials are orthogonal and their support is real,  the only solutions are polynomials of Jacobi, Laguerre or Hermite type.

These orthogonal polynomials are associated with classical random matrix ensembles: Gaussian Unitary Ensemble (Hermite), Laguerre Unitary Ensemble (also known as complex Wishart) and Jacobi Unitary Ensemble (complex MANOVA). Recently, Bornemann~\cite{BORNEMANN},  using the spectral projection method, classified the scaling limits of determinantal processes arising from Sturm-Liouville operators\footnote{He did not used explicitly Bochner theorem.}. They do not lead to any new universality class than what is known for Wishart ensemble.
 
On the other hand, it is known that there are other scaling limits of the kernel in unitary matrix models. These are related to different vanishing of the spectral density at the edge or at the closing gap in the bulk, see~\cite{KuijlaarsUni} for a review. This raises a question whether such limits can be related to spectral projections. To avoid limitations of the Bochner theorem one may look at the class of Hamiltonians with higher powers of momentum operator. Self-adjointness constrains these Hamiltonians to have only  even powers of momentum and Krall~\cite{Krall} provided complete classification of orthogonal polynomials to the problem with quartic momenta. However, classification of higher order Bohner-Krall polynomial systems remains still an open problem. While there are some particular examples of sixth~\cite{Littlejohn1} and eighth order systems~\cite{Azad}, the corresponding weights are only modifications of classical Gaussian, Laguerre and Jacobi weights by Heaviside theta and Dirac delta functions (see~\cite{Littlejohn} for review), which makes them uninteresting from the random matrix theory perspective. 

\subsection{Non-Hermitian `Hamiltonians'}
Relaxing the self-adjointness condition admits a broader class of operators. Then one deals with non-Hermitian `Hamiltonian' and two eigenequations to each eigenvalue:
\be
{\cal H} \ket{P_k}=\lambda_k \ket{P_k}  \,\,\,\,{\rm and} \,\,\,\,\,\, {\cal H}^{\dagger}\ket{Q_k} =\lambda_k \ket{Q_k}. 
\ee 
Here $\ket{P_k}$ and $\bra{Q_k}$ are called left and right eigenfunctions, in the analogy to non-Hermitian matrices. They are no longer orthogonal, but bi-orthogonal
\begin{equation}
\braket{Q_k}{P_l}=\int Q_k(x)P_l(x)dx=\delta_{kl}.
\end{equation}
The adjoint Hamiltonian $\cH^{\dagger}$ is defined in a standard way
 \begin{equation}
 \int f(x)\cH g(x)dx=\int (\cH^{\dagger} f(x))g(x)dx.
 \end{equation}
Now, because of biorthogonality the two sets of eigenfunctions cannot be both polynomials, enlarging the space of possible solutions. 

Preiser~\cite{Preiser} considered a higher order generalization of Bochner-Krall theorem with restriction that $P_k(x)$ are polynomials in $x$, while $Q_k(x)$ are polynomials in $x^m$ multiplied by some weight. He found that for the Hamiltonian with third derivative there exists only one such set, which was discovered earlier by Spencer and Fano~\cite{SpencerFano}. 

Biorthogonal structures appear in multi matrix models, where the correlation kernel is built from biorthogonal functions $P_k$ and $Q_k$
\begin{equation}
K_{N}(x,y)=\sum_{k=0}^{N-1}Q_k(x)P_k(y).\label{eq:KernelBiortho}
\end{equation}

Biorthogonality ensures that the kernel is a projection. It is therefore tempting to ask whether such kernels ale built of eigenfunctions of a certain `Hamiltonian' and if so, is it possible to obtain the microscopic scaling using spectral projections?

\subsection{Singular values of products of complex Gaussian matrices}
Let us consider $X_k$  a rectangular matrices of size $(N+\nu_{k-1})\times(N+\nu_{k})$ with complex Gaussian iid entries of zero mean and unit variance. Without loss of generality we assume $\nu_0=0$ and $\nu_{k}>0$ for $k>0$. The squared singular values of the product $Y_M=X_1X_{2}\ldots X_M$ form a biorthogonal ensemble with the correlation kernel \eqref{eq:KernelBiortho}. The biorthogonal functions are explicitly given by~\cite{AIKWishart}
\begin{eqnarray}
P_k(x)= \MeijerG{1}{0}{1}{M+1}{k+1}{0,-\nu_M,\ldots,-\nu_1}{x}, \\
Q_k(x)=\MeijerG{M}{1}{1}{M+1}{-k}{\nu_M,\ldots,\nu_1,0}{x}. \label{eq:WishQ}
\end{eqnarray} 
Here $G$ stands for the Meijer-G function (see Appendix~\ref{sec:MeijerG}).
From the differential equation \eqref{eq:DiffEqMeijer} we deduce that polynomials $P_k$ satisfy the eigenproblem ($\cH_M P_k=\lambda_k P_k$ with $\lambda_k=k$) of the following differential operator (Hamiltonian)
\begin{equation}
\cH_M=x\frac{d}{dx}-\frac{d}{dx}\prod_{j=1}^{M}\left(x\frac{d}{dx}+\nu_j\right). \label{eq:DiffEqP}
\end{equation}
With the help of the identity $\left(\frac{d}{dx}x-\nu_j\right)\frac{d}{dx}=\frac{d}{dx}\left(x\frac{d}{dx}-\nu_j\right)$ we immediately obtain its adjoint 
\begin{equation}
\cH^{\dagger}_M=-x\frac{d}{dx}-1+(-1)^M\frac{d}{dx}\prod_{j=1}^{M}\left(x\frac{d}{dx}-\nu_j\right).
\end{equation}
The explicit form \eqref{eq:WishQ} and the differential equation \eqref{eq:DiffEqMeijer} prove that $Q_k$ satisfy the eigenequation $\cH^{\dagger}_MQ_k=kQ_k$. Therefore $P_k$ and $Q_k$ are left and right eigenfunctions of a non-Hermitian Hamiltonian.

To probe the microscopic scaling at the edge, we rescale $x=\frac{z}{N}$, which turns the eigenequation for $\cH_M$ into
\begin{equation}
\left[\frac{1}{N}z\frac{d}{dz}-\frac{d}{dz}\prod_{j=1}^M\left(z\frac{d}{dz}+\nu_j\right)\right]P_k=\frac{k}{N}P_k.
\end{equation}
As $k$ is always smaller than $N$, in the large $N$ limit we obtain the condition
\begin{equation}
\Delta_{\vec{\nu}}^{(M+1)}:=-\frac{d}{dz}\prod_{j=1}^M\left(z\frac{d}{dz}+\nu_j\right)\leq 1.
\end{equation}

In order  to continue the analogy to  the deformation of the phase-space of Hermitian operators, we have to find the suitable transformation, which will convert the operator-valued inequality  into an algebraic  constraint.

\subsection{The Narain transform}

In a series of papers~\cite{Narain1,Narain2,Narain3} Narain introduced a broad class of asymmetric transforms, which include many known classical transforms. The Narain transform and its inverse are defined as
\begin{equation}
g(s)=\int_0^{\infty}k(s,y)f(y)dy,\qquad f(y)=\int_0^{\infty}h(y,s) g(s)ds,
\end{equation} 
where the integral kernels read
\begin{eqnarray}
k(s,y)=2\gamma x^{\gamma-1/2}\MeijerG{m}{p}{p+q}{m+n}{a_1,\ldots,a_p,b_1,\ldots,b_q}{c_1,\ldots,c_m,d_1,\ldots,d_n}{(sy)^{2\gamma}}, \\ 
h(y,s)=2\gamma x^{\gamma-1/2}\MeijerG{n}{q}{p+q}{m+n}{-b_1,\ldots,-b_q,-a_1,\ldots,-a_p}{-d_1,\ldots,-d_n,-c_1,\ldots,-c_m}{(ys)^{2\gamma}}.
\end{eqnarray}
If $f$ has a discontinuity at $x$, then $\int_0^{\infty}h(x,s)ds\int_0^{\infty}k(s,y)f(y)dy$ takes the value $\frac{1}{2}(f(x+0)+f(x-0))$, provided that $\sum a_k+\sum b_k=\sum c_k+\sum d_k$.

\subsection{Spectral projection for products of Wishart Matrices}

We use the following kernels in the Narain transformation
\begin{equation}
k(s,y)=\MeijerG{M}{0}{0}{M+1}{-}{\nu_1,\ldots,\nu_M,0}{sy},\qquad h(y,s)=\MeijerG{1}{0}{0}{M+1}{-}{0,-\nu_1,\ldots,-\nu_M}{sy}.
\end{equation}

In the space of the dual variable $s$, the operator $\Delta_{\vec{\nu}}^{(M+1)}$ acts by multiplying by $s$, as can be easily proven, using identities from Appendix~\ref{sec:MeijerG}. 
The hard edge scaling of the kernel reduces therefore the range of parameter $s$ to $s \leq1$. 
Alike in the Hermitian case, the identity operator 
\be
g(x)=\int_0^{\infty}\left[ \int_0^{\infty} h(x,s)k(s,y) ds \right] g(y) dy 
\ee
is  deformed to
\be
{\bf {\rm P}}[g(x)]=\int_0^{\infty}\left[ \int_0^1 h(x,s)k(s,y)ds \right] g(y) dy.
\ee
We obtain this way  the  limiting form of the microscopic kernel at the hard edge
\begin{equation}
K^{hard}_M(x,y)=\int_0^{1}\MeijerG{1}{0}{0}{M+1}{-}{0,-\nu_1,\ldots,-\nu_M}{sx}\MeijerG{M}{0}{0}{M+1}{-}{\nu_1,\ldots,\nu_M,0}{sy}ds.
\end{equation}  
Note that $\MeijerG{1}{0}{0}{2}{-}{\nu,0}{x}=x^{\nu/2}J_{\nu}(2\sqrt{x})$ and $\MeijerG{1}{0}{0}{2}{-}{0,-\nu}{x}=x^{-\nu/2}J_{\nu}(2\sqrt{x})$, which yields
\begin{equation}
K^{hard}_1(x,y)=\left(\frac{y}{x}\right)^{\nu/2}\int_0^{1}J_{\nu}(2\sqrt{sx})J_{\nu}(2\sqrt{sy})ds.
\end{equation}
This form slightly differs from \eqref{eq:BesselKernel}. To understand this discrepancy, let us note that biorthogonal functions can be rescaled as $P_k(x)\to f(x) P_{k}(x)$ and $Q_{k}(x)\to \frac{1}{f(x)}Q_k(x)$ without altering their biorthogonality. Under such a rescaling kernel is transformed $K(x,y)\to \frac{1}{f(x)}K(x,y)f(y)$. In our case it is sufficient to take $f(x)=x^{\nu/2}$ and further rescale $(x,y)\to \frac{1}{4}(x,y)$. 
The Narain transform can therefore be viewed as a generalization of the Hankel transform at the hard edge.

\subsection{Muttalib-Borodin ensemble with the Laguerre weight}

As another example we consider the joint pdf of eigenvalues introduced by Muttalib~\cite{Muttalib} and elaborated later by Borodin~\cite{BorodinBiOrtho}
\begin{equation}
P(\lambda_1,\ldots,\lambda_n)=C_N \prod_{1\leq i<j\leq N}|\lambda_i-\lambda_j| \prod_{1\leq i<j\leq N}|\lambda_i^{\theta}-\lambda_j^{\theta}| \prod_{k=1}^{N}\lambda_k^\alpha e^{-\lambda_k} d\lambda_k,
\end{equation}
with $\alpha>-1$, and $\theta\geq 0$.
Eigenvalues form a determinantal point process with a correlation kernel given by the bi-orthogonal functions~\eqref{eq:KernelBiortho}.
Here $P_k$ is a polynomial of order $k$, while $Q_k$ is a polynomial in $x^\theta$ multiplied by the Laguerre weight. For integer values of $\theta$ Konhauser provides the explicit form of Q~\cite[eq. (5)]{Konhauser}
\begin{eqnarray}
Q_k(x)=x^{\alpha}e^{-x}\sum_{j=0}^{k}(-1)^j {k\choose j}\frac{x^{j\theta}}{\Gamma(j\theta+\alpha+1)},
\end{eqnarray}
while Carlitz gives the explicit form of polynomials~\cite[eq. (9)]{Carlitz}
\begin{equation}
P_k(x)=\frac{1}{k!}\sum_{i=0}^{k}\frac{x^i}{i!}\sum_{j=0}^i(-1)^j{i \choose j} \frac{\Gamma(k+\frac{j+\alpha+1}{\theta})}{\Gamma(k)}.
\end{equation}
For $\theta=1$ this reduces to the Laguerre orthogonal polynomials, while the case $\theta=2$ was considered by Preiser~\cite{Preiser} in an attempt to extend Bohner-Krall theorem. Polynomials satisfy the eigenvalue equation $\cH P_k=\lambda_k P_k$, with $\lambda_k=\theta k$ of the following differential operator~\cite{Konhauser}
\begin{equation}
\cH=\left(\frac{d}{dx}x+\alpha-x\right)\left[\left(1-\frac{d}{dx}\right)^{\theta}-1\right].
\end{equation}
Konhauser showed also that $Z_k=x^{-\alpha}e^{x}Q_k(x)$, a polynomial in $x^\theta$, satisfies~\cite[eq. (10)]{Konhauser}
\begin{equation}
\left(\frac{d}{dx}\right)^\theta x^{\alpha+1}\frac{d}{dx}Z_k-x^{\alpha+1}\frac{d}{dx}Z_k=-x^{\alpha}\theta k Z_k.
\end{equation}
Then it is straightforward to show that $Q_k$ satisfies the eigenequation $\cH^{\dagger}Q_k=\lambda_kQ_k$ to the same eigenvalues as $P_k$. The differential operator
\begin{equation}
\cH^{\dagger}=\left[1-\left(1+\frac{d}{dx}\right)^{\theta}\right]\left(-\alpha+x+x\frac{d}{dx}\right)
\end{equation}
is the adjoint of $\cH$.
We probe the hard edge by introducing a new variable $x=u N^{-\frac{1}{\theta}}$. 
In the large $N$ limit, having in mind that $k<N$, from the eigenequation for $\cH^{\dagger}$ we obtain
\begin{equation}
-\frac{1}{\theta}\left(\frac{d}{du}\right)^{\theta}\left(u\frac{d}{du}-\alpha\right) \leq 1.
\end{equation}
A change of variable  $u=\theta z^{1/\theta}$ turns this conditions into a more familiar form
\begin{eqnarray}
-\frac{d}{dz}\prod_{j=1}^{\theta}\left(z\frac{d}{dz}+\nu_j\right)\leq 1, \label{eq:MBcond}
\end{eqnarray}
with 
\begin{equation}
\nu=-\frac{1}{\theta},-\frac{2}{\theta},-\frac{3}{\theta},\ldots,-\frac{\theta-1}{\theta},-\frac{\alpha}{\theta}.
\end{equation}
We now take
\begin{equation}
k(s,y)=\MeijerG{\theta}{0}{0}{\theta+1}{-}{0,-\frac{1}{\theta},\ldots,-\frac{\theta-1}{\theta},-\frac{\alpha}{\theta}}{sy},\qquad h(y,s)=\MeijerG{1}{0}{0}{\theta+1}{-}{\frac{\alpha}{\theta},0,\frac{1}{\theta},\ldots,\frac{\theta-1}{\theta}}{sy}.
\end{equation} 
Again, using the identities from Appendix~\ref{sec:MeijerG} one can show that $\int_0^{\infty}k(s,z)(\cH^{\dagger} f(z))dz=\int_0^{\infty}sk(s,z)f(z)dz$. This means that the condition \eqref{eq:MBcond} in the dual space is eqivalent to $s\leq 1$. This allows us to read out the form of the kernel
\begin{equation}
K(y,x)=\int_0^{1}\MeijerG{1}{0}{0}{\theta+1}{-}{\frac{\alpha}{\theta},0,\frac{1}{\theta},\ldots,\frac{\theta-1}{\theta}}{sx}\MeijerG{\theta}{0}{0}{\theta+1}{-}{0,-\frac{1}{\theta},\ldots,-\frac{\theta-1}{\theta},-\frac{\alpha}{\theta}}{sy} ds.
\end{equation}
Note also that the truncation condition $s\leq 1$ was obtained from the consideration of $\cH^{\dagger}$, therefore the kernel has now interchanged arguments. Using \eqref{eq:MeijerAbsorb} we also write an equivalent kernel
\begin{equation}
\left(\frac{y}{x}\right)^{\frac{\alpha}{\theta}} K(y,x)=\int_0^1\MeijerG{1}{0}{0}{\theta+1}{-}{0,-\frac{\alpha}{\theta},-\frac{\alpha-1}{\theta},\ldots,-\frac{\alpha-\theta+1}{\theta}}{sx}
\MeijerG{\theta}{0}{0}{\theta+1}{-}{\frac{\alpha}{\theta},\frac{\alpha-1}{\theta},\ldots,\frac{\alpha-\theta+1}{\theta},0}{sy}ds,
\end{equation}
which corresponds to the form obtained by Kuijlaars and Stivigny~\cite[Theorem 5.1]{KuijStiv}.

\section{Summary}
We start from historical digression. 
It is intriguing to investigate   the chronological intertwining of the  ideas in quantum mechanics, mathematics  and statistics from the perspective of the contemporary random matrix theory. 
In 1926, Schr\"odinger has solved  his equation for Coulomb potential, obtaining among others the radial parts of the wave  function in terms of Laguerre functions\footnote{Year earlier, Pauli has quantized algebraically hydrogen atom, using the hidden symmetry (Runge-Lenz vector) of the Coulomb potential, therefore treating  this problem as a free problem on $S_3$ hypersphere.}. 
Two years later (1928) Wishart introduced his ensemble in multivariate statistics,  as a generalization of the $\chi^2$ ensemble~\cite{WISHART}. The original paper deals with the real random variables, but his ideas were later generalized to complex  variables~\cite{Goodman1963}.   A year later (1929),  Bochner has proven his theorem~\cite{Bochner} for Sturm--Liouville operators, without  any direct references to Schr\"odinger equation.  At that time spectral properties of random matrices were not considered at all.   Laguerre  polynomials  appeared explicitly in random matrix theory  only after the Mehta and Gaudin used the orthogonal polynomial trick to disentangle the Van der Monde determinant~\cite{Mehta1960}.   This technique has also  paved the way for classical universal kernels. However, the link to the uniqueness of the determinantal  triality of soft, edge and bulk microscopic universalities of Sturm--Liouville operators have been cleared out only recently~\cite{BORNEMANN}.

In 1967 Marchenko and Pastur  derived the spectral density for the Wishart ensemble~\cite{Marcenko1967}. Interestingly, they used the ideas borrowed from hydrodynamics~\cite{PasturPriv}. The fact that the Marchenko-Pastur distribution can be interpreted as a Bohr-Sommerfeld quantization condition for the hydrogen atom was not, to the best of or knowledge,  explicitly stated in the literature. Such a link is intuitively expected, because the Dyson electrostatic analogy in the limit of large matrices  allows one to solve the random matrix model using the saddle point approximation - the same mathematical method which gives the WKB approximation in Quantum Mechanics, with the correspondence  $\frac{1}{\hbar}  \leftrightarrow N$. The relation between momentum and the spectral density  requires, however,  additional  knowledge of the properties of Wigner functions, as we  point out in this paper. 
 
With introducing non-trivial initial conditions for Dyson Brownian motion,  new universality classes emerged in random matrix theory.  In the 90's of the previous century,  collision of soft  edges in GUE led Brezin and Hikami~\cite{BreHi} to the Pearcey  kernel. In a similar collision of chiral fronts at the hard edge of the chiral random matrix model one of the authors found the Bessoid  kernel universality~\cite{Bessoid}. While still determinantal~\cite{ZinnJustin1997}, such models break rotational invariance, and require non-standard tools. Later it was discovered that such ensembles can be solved by polynomials that are orthogonal to more than one weight~\cite{Bleher2004}.

The bi-orthogonality method of Muttalib and Borodin opened a new way for treating a broader class of random matrix models, to which the orthogonal polynomials method does not apply.  Historically, it is again puzzling  that  bi-orthogonality was not linked to random matrices earlier. Already in 1951, Fano and Spencer~\cite{SpencerFano} studying  propagation of the X-rays through the matter, have introduced bi-orthogonal Laguerre polynomials.  These ideas  were further developed in mathematics by Preiser~\cite{Preiser} and Konhauser~\cite{Konhauser}. In particular, Preiser's construction corresponds  exactly to the case of Muttalib-Borodin ensemble.
 
 This is precisely that intertwining of ideas and the lack of explicit ideas, which prompted us to reexamine Bochner theorem. Rapid progress in random matrix theory in last three decades has brought plethora of new microscopic universality classes. Despite so many examples of microscopic universalities, there is lack of their systematic classification. The spectral projection method adopted to non-Hermitian Hamiltonians and possible generalizations of Bochner theorem for higher order differential operators\footnote{After completion of the paper, Oleg Evnin has pointed to us a recent paper~\cite{HOROZOV}, proposing the classification of cubic extension of Bochner theorem.  This construction, however, does not refer to random matrix theory.} offer a new perspective on this problem. Certainly, this program is a challenging  mathematical problem, which we do not attempt to solve.

This work raises a series of fundamental questions related to possible generalization of Bochner theorem in the context of random matrix theory. Is it possible to reframe all universality classes in this language? Will this classification be predictive for constructing new types of random matrix models?  Can one  infer the microscopic kernels of non-Hermitian ensembles from a `complex version' of Bochner theorem?  We leave these questions open but we think that the presented method has also pedagogical value. It offers an easy  and intuitive way to recover not only the the classical universality classes, but also  more involving Meijer-G functions. Combining physical intuition with  mathematics may provide in such a way new insights even in standard problems.

\section*{Acknowledgements}

  The research was supported by the MAESTRO DEC-2011/02/A/ST1/00119 grant of the National Center of Science. WT also appreciates the financial support from the Polish Ministry of Science and Higher Education through ``Diamond Grant" 0225/DIA/2015/44 and the doctoral scholarship ETIUDA UMO-2018/28/T/ST1/00470 from National Science Center.

\begin{appendices}
\section{Mapping Wishart onto 2D hydrogen atom} \label{sec:Hydro2D}

The time-independent Schr\"odinger equation in 2D with the potential $V(r)=-Ze^2/r$ in the cylindrical coordinates reads
\begin{equation}
\left[-\frac{\hbar^2}{2m}\left(\frac{\partial^2}{\partial r^2}+\frac{1}{r}\frac{\partial}{\partial r}+\frac{1}{r^2}\frac{\partial^2}{\partial\varphi^2}\right)-Ze^2/r\right]\phi(r,\varphi)=E\varphi(r,\varphi).
\end{equation}
An Ansatz $\phi(r,\varphi)=R(r)e^{il\varphi}/\sqrt{2\pi}$ separates variables. Setting the physical constants $Ze^2=1$, $2m=1$, $\hbar=1$ and changing variables as $\rho=\lambda r$, $E=-1/4\lambda^2$ we arrive at the equation for the radial part
\begin{equation}
\left(\frac{d^2}{d\rho^2}+\frac{1}{\rho}\frac{d}{d\rho}+\frac{\lambda}{\rho}-\frac{l^2}{\rho^2}-\frac{1}{4}\right) R(\rho)=0.
\end{equation}
Upon identification $2l=|\alpha|$ and $2\lambda=1+2k+\alpha$ we obtain the equation \eqref{mainequation_no1} for the function building the kernel.

\section{WKB analysis of the macroscopic spectral density} \label{sec:AppWKB}

The spectral density is calculated from the kernel as
\begin{equation}
\rho(x)=\frac{1}{N}K(x,x)=\frac{1}{N}\sum_{k=0}^{N-1}\psi_k^2(x).
\end{equation}
In the large $N$ limit the sum can be approximated by an integral over the variable $t=k/N$
\begin{equation}
\rho(x)\xrightarrow{N\to\infty} \int_0^1\psi_t^2(x)dt.
\end{equation}
Taking the equation \eqref{mainequation_no1} for $\psi_k$, rescaling $x\to Tx$ and setting $t=k/N$, we obtain
\begin{equation}
\frac{1}{T^2}\left(\frac{d^2}{dx^2}+\frac{1}{x}\frac{d}{dx}\right)\psi_t(x)=\left(\frac{1}{4}+\frac{(1-c)^2}{4x^2}-\frac{ct}{x}-\frac{1-c}{2x}\right)\psi_t(x)\equiv (V(x)-E)\psi_t(x). \label{eq:WKB}
\end{equation}
We also note that up to a term $1/4x^2$, which is irrelevant in the asymptotic analysis, the operator on the lhs of \eqref{eq:WKB} is minus square of the radial momentum $p_{r}(x)=-i\hbar \left(\frac{1}{r}+\frac{d}{dr}\right)$. 
Using the WKB Ansatz $\psi(x)=A(x)e^{T\phi(x)}$, we obtain
the general solution
\begin{equation}
\psi_t(x)=\frac{1}{\sqrt{x p_r(x)}}\left(C_{+}e^{iT\int^x p_r(x')dx'}+C_{-}e^{-iT\int^xp_r(x')dx'}\right).
\end{equation}
Matching condition at each of the turning points gives two forms of the solution
\begin{equation}
\psi(x)=\frac{C}{\sqrt{x p_r(x)}}\cos\left[-\frac{\pi}{4}+T\int_{x_{-}}^{x}dx'p_r(x')\right]=
\frac{C'}{\sqrt{x p_r(x)}}\cos\left[-\frac{\pi}{4}+T\int_{x}^{x_{+}}dx'p(x')\right].
\end{equation}
Uniqueness of the solution irrespective of the choice of turning point leads to the quantization condition
\begin{equation}
T\oint p_r(x)dx=2\pi\left(n+\frac{1}{2}\right),\quad n\in\mathbb{N}.
\end{equation}
Note that for the calculation of the spectral density, $\psi_t^2$ is needed. For large $T$ it is a rapidly oscillating function and the oscillations average out and only the average of $\cos^2$, which is $1/2$, is relevant\footnote{This can be rephrased more rigorously in terms of weak convergence.} 
\begin{equation}
\psi^2_t(x)=\left\lbrace\begin{array}{ccc}
0 & \mbox{for} & x<x_{-} \mbox{ or } x>x_{+} \\
\frac{C}{2x p(x,t)} & \mbox{for}  & x_{-}<x<x_{+}
\end{array}\right..
\end{equation}
The turning points are
\begin{equation}
x_{\pm}(t)=1-c+2ct\pm 2\sqrt{ct(1+ct-c)}.
\end{equation}
The spectral density is therefore given by
\begin{equation}
\rho(x)=\int_0^1 dt \frac{C}{\sqrt{2c(1+2tx-x)-c^2-(x-1)^2}}\chi_{x_{-}<x<x_{+}}=\frac{C}{2cx}\sqrt{(x-(1-\sqrt{c})^2)((1+\sqrt{c})^2-x)},
\end{equation}
where $\chi_A$ is equal to 1 when $A$ is true and 0 for $A$ false. Setting $C=\frac{1}{\pi}$ normalizes the density.

\section{Some properties of Meijer-G functions} \label{sec:MeijerG}
The Meijer-G functions are defines as an integral
\begin{equation}
\MeijerG{m}{n}{p}{q}{a_1,\ldots,a_p}{b_1,\ldots,b_q}{z}=\frac{1}{2\pi i}\int_L\frac{ \prod_{j=1}^{m}\Gamma(b_j-s)\prod_{j=1}^{n}\Gamma(1-a_j+s)}{\prod_{j=m+1}^{q}\Gamma(1-b_j+s)\prod_{j=n+1}^p \Gamma(a_j-s)}z^s ds,
\end{equation}
where $\Gamma(z)$ is the Euler gamma function. The integration contour $L$ is chosen to separate all poles of $\prod_{j=1}^{m}\Gamma(b_j-s)$ from the poles of $\prod_{j=1}^{n}\Gamma(1-a_j+s)$ (see also \cite{Luke}, \textsection 5.2 for details). By definition, they are symmetric in its first $m$ and last $q-m$ lower parameters. When first and the last lower parameter differ by an integer number, they can be interchanged
\begin{equation}
\MeijerG{m}{n}{p}{q}{a_1,\ldots,a_p}{b_1,b_2,\ldots,b_{q-1},b_q}{z}=(-1)^{b_q-b_1}\MeijerG{m}{n}{p}{q}{a_1,\ldots,a_p}{b_q,b_2,\ldots,b_{q-1},b_1}{z}. \label{eq:MeijerInterchange}
\end{equation}
The following differential operator acts by increasing first lower indices
\begin{equation}
\left( -z\frac{d}{dz}+b_1\right)\MeijerG{m}{n}{p}{q}{a_1,\ldots, a_p}{b_1,b_2,\ldots,b_q}{z}=\MeijerG{m}{n}{p}{q}{a_1,\ldots, a_p}{b_1+1,b_2,\ldots,b_q}{z}. \label{eq:DiffMeijer1}
\end{equation}
Combining this with \eqref{eq:MeijerInterchange}, we obtain the operator $z\frac{d}{dz}-b_q$, which increases last lower indices.
Multiplication by the argument allows one to increase all indices
\begin{equation}
z^{\alpha}\MeijerG{m}{n}{p}{q}{a_1,\ldots,a_p}{b_1,\ldots,b_q}{z}=\MeijerG{m}{n}{p}{q}{a_1+\alpha,\ldots,a_p+\alpha}{b_1+\alpha,\ldots,b_q+\alpha}{z}. \label{eq:MeijerAbsorb}
\end{equation}
Meijer-G functions satisfy the following differential equation
\begin{equation}
\left[(-1)^{p-m-n}z\prod_{j=1}^{p}\left(z\frac{d}{dz}-a_j+1\right)-\prod_{j=1}^{q}\left(z\frac{d}{dz}-b_j\right)\right]\MeijerG{m}{n}{p}{q}{a_1,\ldots,a_p}{b_1,\ldots,b_q}{z}=0. \label{eq:DiffEqMeijer}
\end{equation}

\end{appendices}

\bibliography{bibliography}

\begin{thebibliography}{57}
\bibitem [{NOT()}]{NOTE}%
  \BibitemOpen
  \href@noop {} {}\bibinfo {note} {Simple exercise shows that {WKB}
  approximation yields the formula $E_{WKB}=E_0/[(n-1/2)^2+\sqrt{l(l+1)}]^2$,
  where $E_0=-13.6 eV$}\BibitemShut {NoStop}%
\end{thebibliography}

\end{document}